\documentclass[pre,twocolumn,aps,eqsecnum]{revtex4}
\usepackage{amssymb,amstext,amsmath}
\usepackage{graphicx}
\begin{document}

\title{Thermodynamic dislocation theory: torsion of bars} 
\author{K.C. Le$^{1,2}$, Y. Piao$^3$, T.M. Tran$^{1,2}$} 
\affiliation{$^1$\,Materials Mechanics Research Group, Ton Duc Thang University, Ho Chi Minh City, Vietnam\\
$^2$\,Faculty of Civil Engineering, Ton Duc Thang University, Ho Chi Minh City, Vietnam\\
$^3$\,Lehrstuhl f\"ur Mechanik - Materialtheorie, Ruhr-Universit\"at Bochum, D-44780 Bochum, Germany}

\date{\today}

\begin{abstract}
The thermodynamic dislocation theory developed for non-uniform plastic deformations is used here in an analysis of a bar subjected to torsion. Employing a small set of physics-based parameters, which we expect to be approximately independent of strain rate and temperature, we are able to simulate the torque-twist curve for a bar made of single crystal copper that agrees with the experimental one.  
\end{abstract}

\maketitle

\section{Introduction}
\label{Intro} 

The thermodynamic dislocation theory (TDT), proposed initially by Langer, Bouchbinder, and Lookman \cite{LBL-10} and developed further in \cite{JSL-15,JSL-16,JSL-17,JSL-17a,Le17,Le18}, deals with the uniform plastic deformations of crystals driven by a constant strain rate. During these uniform plastic deformations the crystal may have only redundant dislocations whose resultant Burgers vector vanishes. As shown in \cite{Le18a,LP18,LeTr18}, the extension of TDT to non-uniform plastic deformations should account for excess dislocations due to the incompatibility of the plastic distortion \cite{Nye53}. There are various examples of non-uniform plastic deformations in material science and engineering, the most typical of which being the torsion of bars \cite{Fleck94} and the bending of beams \cite{Stoelken98}. The purpose of this paper is to explore use of TDT for non-uniform plastic deformations \cite{Le18a,LP18,LeTr18} in modeling bars made of single crystal copper and subjected to torsion. Our challenge is to simulate the torque-twist curve exhibiting the hardening behavior and the size effect. We also want to compare this torque-twist curve with the experimental curve provided by Horstemeyer et al. \cite{Horstemeyer02}. To make this comparison possible we will need to identify from the experimental data obtained in \cite{Horstemeyer02} a list of material parameters for single crystal copper under torsion.  For this purpose, we will use the large scale least-squares analysis described in \cite{Le17,Le18,LeTr17}. 

The thermodynamic dislocation theory is based on two unconventional ideas. The first of these is that, under nonequilibrium conditions, the atomically slow configurational degrees of freedom of dislocated crystals are characterized by an effective disorder temperature that differs from the ordinary kinetic-vibrational temperature. Both of these temperatures are thermodynamically well defined variables whose equations of motion determine the irreversible behaviors of these systems. The second principal idea is that entanglement of dislocations is the overwhelmingly dominant cause of resistance to deformation in crystals.  These two ideas have led to successfully predictive theories of strain hardening \cite{LBL-10,JSL-15}, steady-state stresses over exceedingly wide ranges of strain rates \cite{LBL-10}, thermal softening during deformation \cite{Le17}, yielding transitions between elastic and plastic responses  \cite{JSL-16,JSL-17a}, shear banding instabilities \cite{JSL-17,Le18}, and size and Bauschinger effects \cite{Le18a,LP18,LeTr18}.  

We start in Sec.~\ref{EOM} with a brief annotated summary of the equations of motion to be used here.  Our focus is on the physical significance of the various parameters that occur in them.  We discuss which of these parameters are expected to be material-specific constants, independent of temperature and strain rate, and thus to be key ingredients of the theory. In Sec.~\ref{NI} we discretize the obtained system of governing equations and develop the numerical method for its solution. The parameter identification based on the large scale least squares analysis and the results of the numerical simulations are presented in Sec.~\ref{PI}.  We conclude in Sec.\ref{CONCLUSIONS} with some remarks about the significance of these calculations.
 
\section{Equations of Motion}
\label{EOM}
 
Suppose a single crystal bar with a circular cross section, of radius $R$ and length $L$, is subjected to torsion (see drawing of the bar with its cross-section in Fig.~1). For this particular geometry of the bar and under the condition $R\ll L$ it is natural to assume that the warping of the bar vanishes, while the circumferential displacement is $u_\varphi=\omega rz$, with $\omega $ being the twist angle per unit length. Thus, the total shear strain of the bar $\gamma =2\epsilon_{\varphi z}=\omega r$ and the shear strain rate $\dot{\gamma}=\dot{\omega}r$ turn out to be non-uniform as they are  linear functions of radius $r$. 

\begin{figure}[t]
	\centering
	\includegraphics[width=.4\textwidth]{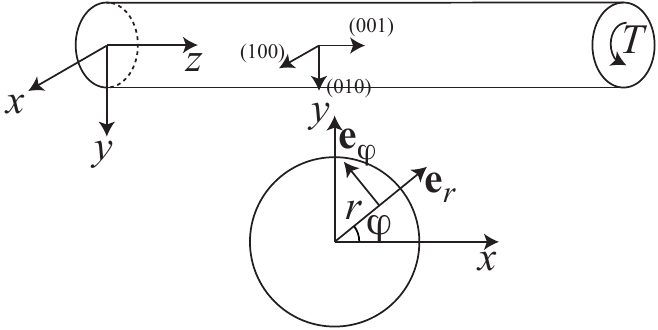}
	\caption{Torsion of a single crystal bar}
	\label{bar}
\end{figure}

Now, let this system be driven at a constant twist rate $\dot\omega \equiv \varpi_0/t_0$, where $t_0$ is a characteristic microscopic time scale. Since the system experiences a steady state torsional deformation, we can replace the time $t$ by the total twist angle (per unit length) $\omega$ so that $t_0\,\partial/\partial t \to \varpi_0\,\partial/\partial \omega$. The equation of motion for the flow stress becomes
\begin{equation}
\label{tauydot}
\frac{\partial \tau_Y}{\partial \omega} = \mu\,\left[r - \frac{q(\gamma )}{\varpi_0}\right],
\end{equation}
with $\mu$ being the shear modulus. This equation is derived from Eq. (II.1) in \cite{LeTr18} by replacing $\gamma =r \omega$ and multiplying both sides by $r$. Note that for uniform plastic deformations involving only redundant dislocations $q(\gamma )/t_0$ equals the plastic shear rate $\dot{\beta}$, with $\beta $ being the uniform plastic distortion. However, if $\beta $ is non-uniform, it is not necessarily so.

The state variables that describe this system are the elastic strain $\gamma-\beta$, the areal densities of redundant dislocations $\rho_r$ and excess dislocations $\rho_g\equiv  |\beta _{,r}+\beta/r|/ b$ (where $b$ is the length of the Burgers vector), and the effective disorder temperature $\chi$ (cf. \cite{Kroener1992,JSL-16}). All four quantities, $\gamma-\beta$, $\rho_r$, $\rho_g$, and $\chi$, are functions of $r$ and $\omega$. 

The central, dislocation-specific ingredient of this analysis is the thermally activated depinning formula for $q$ as a function of a flow stress $\tau_Y$ and a total dislocation density $\rho=\rho_r+\rho_g$:  
\begin{align}
\label{qdef}
q(\tau_Y,\rho)&= b\sqrt{\rho} [f_P(\tau_Y,\rho)-f_P(-\tau_Y,\rho)], 
\\
f_P(\tau_Y,\rho)&=\exp\,\Bigl[-\,\frac{1}{\theta}\,e^{-\tau_Y/\tau_T(\rho)}\Bigr]. \notag
\end{align}
This is an Orowan relation of the form $q = \rho\,b\,v\,t_0$ in which the speed of the dislocations $v$ is given by the distance between them multiplied by the rate at which they are depinned from each other. That rate is approximated here by the activation terms $f_P(\tau_Y,\rho)$ and $-f_P(-\tau_Y,\rho)$, in which the energy barrier $e_P=k_BT_P$ (implicit in the scaling of $\theta=T/T_P$) is reduced by the stress dependent factor $e^{-\tau_Y/\tau_T(\rho)}$, where  $\tau_T(\rho)= \mu_T\,b \sqrt{\rho}$ is the Taylor stress with $\mu_T$ being proportional to $\mu $ (see Section \ref{NI}). Note that antisymmetry is required in Eq.~(\ref{qdef}), especially when dealing with the load reversal, both to preserve reflection symmetry, and to satisfy the second-law requirement that the energy dissipation rate, $\tau_Yq/(\varpi_0r)$, is non-negative.  

The pinning energy $e_P$ is large, of the order of electron volts, so that $\theta$ is very small.  As a result, $q(\tau_Y,\rho)$ is an extremely rapidly varying function of $\tau_Y$ and $\theta$.  This strongly nonlinear behavior is the key to understanding yielding transitions and shear banding as well as many other important features of crystal plasticity.  For example, the extremely slow variation of the steady-state flow stress as a function of strain rate discussed in \cite{LBL-10} is the converse of the extremely rapid variation of $q$ as a function of $\tau_Y$ in Eq.(\ref{qdef}).  

The equation of motion for the total dislocation density $\rho=\rho_r+\rho_g$ describes energy flow. It says that some fraction of the power delivered to the system by external driving is converted into the energy of dislocations, and that that energy is dissipated according to a detailed-balance analysis involving the effective temperature $\chi$. In terms of the twist angle $\omega$ this equation reads: 
\begin{equation}
\label{rhodot}
\frac{\partial \rho}{\partial \omega} = K_\rho \,\frac{\tau_Y\,q}{a^2\nu(\theta,\rho,\varpi_0r)^2\,\mu\,\varpi_0}\, \Bigl[1 -\frac{\rho}{\rho_{ss}(\chi)} \Bigr],
\end{equation}
with $\rho_{ss}(\chi) =(1/a^2)e^{- e_D/\chi}$ being the steady-state value of $\rho$ at given $\chi$, $e_D$ a characteristic formation energy for dislocations, and $a$ denoting the average spacing between dislocations in the limit of infinite  $\chi$ ($a$ is a length of the order of tens of atomic spacings). The coefficient $K_\rho $ is an energy conversion factor that, according to arguments presented in  \cite{LBL-10} and \cite{JSL-17}, should be independent of both strain rate and temperature. The other quantity that appears in the prefactor in Eq.(\ref{rhodot}) is
\begin{equation}
\label{nudef}
\nu(\theta,\rho,q_0) \equiv \ln\Bigl(\frac{1}{\theta}\Bigr) - \ln\Bigl[\ln\Bigl(\frac{b\sqrt{\rho}}{q_0}\Bigr)\Bigr].
\end{equation}

The equation of motion for the effective temperature $\chi$ is a statement of the first law of thermodynamics for the configurational subsystem: 
\begin{equation}
\label{chidot}
\frac{\partial \chi }{\partial \omega} = K\,\frac{\tau_Y e_D \,q}{\mu\,\varpi_0}\,\Bigl( 1 -\frac{\chi}{\chi_0} \Bigr). 
\end{equation}
Here, $\chi_0$ is the steady-state value of $\chi$ for strain rates appreciably smaller than inverse atomic relaxation times, i.e. much smaller than $t_0^{-1}$. The dimensionless factor $K$ is inversely proportional to the effective specific heat $c_{e\!f\!f}$. Unlike $K_\rho$, there is no  reason to believe that $K$ is a rate-independent constant.  In \cite{JSL-17a}, $K$ for copper was found to decrease from $17$ to $12$ when the strain rate increased by a factor of $10^6$.  Since the maximum strain rate (reached at the outer radius of the bar) for the small twist rate in our torsion test is small, we assume that $K$ is a constant.  

The equation for the plastic distortion $\beta$ reads 
\begin{equation}
\label{microforces}
\tau -\tau_B -\tau_Y =0.
\end{equation} 
This equation is the balance of microforces acting on excess dislocations. Here, the first term $\tau=\mu(\gamma -\beta)=\mu(\omega r-\beta)$ is the applied shear stress, the second term the back-stress due to the interaction of excess dislocations, and the last one the flow stress. This balance of microforces can be derived from the variational equation for irreversible processes \cite{Le18a,LP18} yielding
\begin{equation}
\label{zeta}
\tau_B =-\frac{1}{b^2}\frac{\partial ^2\psi_m}{\partial (\rho_g)^2}(\beta_{,rr}+\beta_{,r}-\beta/r^2),
\end{equation}
with $\psi_m$ being the free energy density of excess dislocations. Note that the applied shear stress is equal to the flow stress for the uniform plastic deformations. Berdichevsky \cite{VB17} has found $\psi_m$ for the locally periodic arrangement of excess screw dislocations in a bar under torsion. However, as shown by us in \cite{LP18}, his expression must be extrapolated to the extremely small or large dislocation densities to guarantee the existence of solution within TDT. Using the extrapolated energy proposed in \cite{LP18} we find that $\tau_B$ is given by
\begin{equation}
\label{backstress}
-\mu b^2\frac{k_1\xi^2+(2k_0k_1-1)\xi+k_1k_0^2-2k_0}{4\pi(k_0+\xi)^2}(\beta_{,rr}+\beta_{,r}-\beta/r^2),
\end{equation}
where $\xi=b|\beta_{,r}+\beta/r|$. Equation (\ref{microforces}) is subjected to the boundary conditions $\beta(0)=0$ and $\beta_{,r}(R)+\beta (R)/R=0$. The second condition means that the density of excess dislocations must vanish at the free boundary.     

\begin{figure}[t]
	\centering
	\includegraphics[width=.45\textwidth]{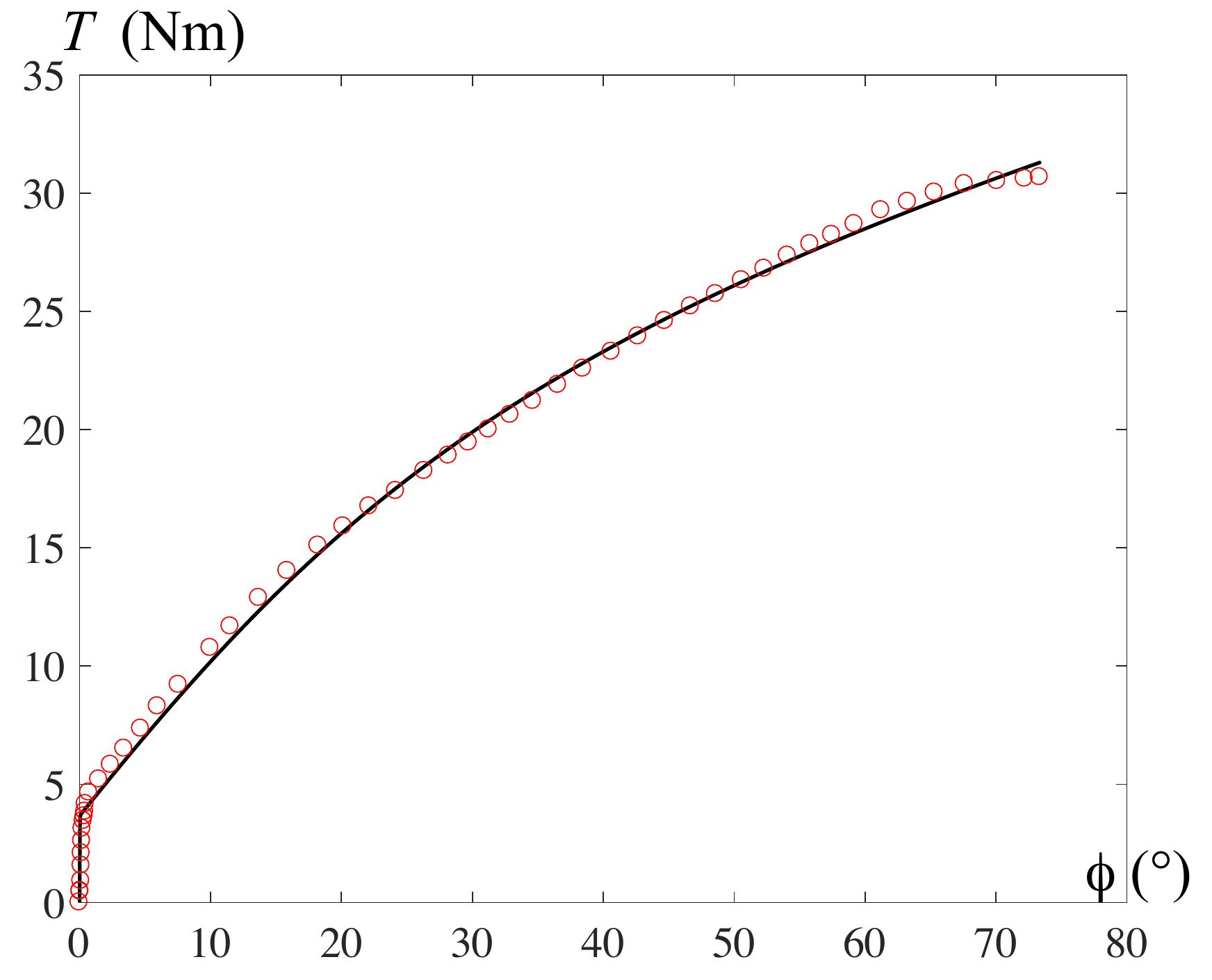}
	\caption{(Color online) The torque-twist curve at the twist rate $\dot{\phi}=0.25^\circ/$s and for room temperature. The experimental points are taken from Horstemeyer et al. \cite{Horstemeyer02}}
	\label{torquetwist}
\end{figure}

\section{Discretization and method of solution}
\label{NI}
For the purpose of numerical integration of the system of equations (\ref{tauydot})-(\ref{backstress}) let us introduce the following variables and quantities
\begin{eqnarray}
\tilde{r}=r/b,\, \tilde{\tau}=\tau/\mu, \, \tilde{\tau}_Y=\tau_Y/\mu ,\, \tilde{\tau}_B=\tau_B/\mu , \notag
\\
\phi=\omega/\eta , \, \eta=\frac{\pi}{180^\circ L},\, \tilde{\rho}=a^2\rho .\label{dimless}
\end{eqnarray}
The variable $\tilde{r}$ changes from zero to $\tilde{R}=R/b$. The variable $\phi$ has the meaning of the total twist angle measured in degree (in \cite{Horstemeyer02} $\phi$ changes from zero to $\phi_*=73.35^\circ$). The calculation of the torque as function of $\phi$ is convenient for the later comparison with the torque-twist curve from \cite{Horstemeyer02}. Then we rewrite Eq.~(\ref{qdef}) in the form
\begin{equation}
\label{tildeq}
q(\tau_Y,\rho)=\frac{b}{a}\tilde{q}(\tilde{\tau}_Y,\tilde{\rho}),
\end{equation}
where
\begin{equation}
\label{tildeqdef}
\tilde{q}(\tilde{\tau}_Y,\tilde{\rho})=\sqrt{\tilde{\rho}}[\tilde{f}_P(\tilde{\tau}_Y,\tilde{\rho})-\tilde{f}_P(-\tilde{\tau}_Y,\tilde{\rho})].
\end{equation}
We set $\tilde{\mu}_T=(b/a)\mu_T=\mu s$ and assume that $s$ is independent of temperature and strain rate. Then
\begin{equation}
\label{tildefp}
\tilde{f}_P(\tilde{\tau}_Y,\tilde{\rho})=\exp\,\Bigl[-\,\frac{1}{\theta}\,e^{-\tilde{\tau}_Y/(s\sqrt{\tilde{\rho }})}\Bigr].
\end{equation}
We define $\tilde{\varpi}_0=(a/b)\varpi_0$ so that $q/\varpi_0=\tilde{q}/\tilde{\varpi}_0$. Eq.~(\ref{nudef}) becomes
\begin{equation}
\label{nudef1}
\tilde{\nu}(\theta,\tilde{\rho},\tilde{\varpi}_0r) \equiv \ln\Bigl(\frac{1}{\theta}\Bigr) - \ln\Bigl[\ln\Bigl(\frac{\sqrt{\tilde{\rho}}}{\tilde{\varpi}_0r}\Bigr)\Bigr].
\end{equation}
The dimensionless steady-state quantities are
\begin{equation}
\label{ss}
\tilde{\rho}_{ss}(\tilde{\chi})=e^{-1/\tilde{\chi}}, \quad \tilde{\chi}_0=\chi_0/e_D.
\end{equation}
Using $\tilde{q}$ instead of $q$ as the dimensionless measure of plastic
strain rate means that we are effectively rescaling $t_0$ by a
factor $b/a$. For purposes of this analysis, we assume that $(a/b)t_0=10^{-12}$s.

\begin{figure}[t]
	\centering
	\includegraphics[width=.45\textwidth]{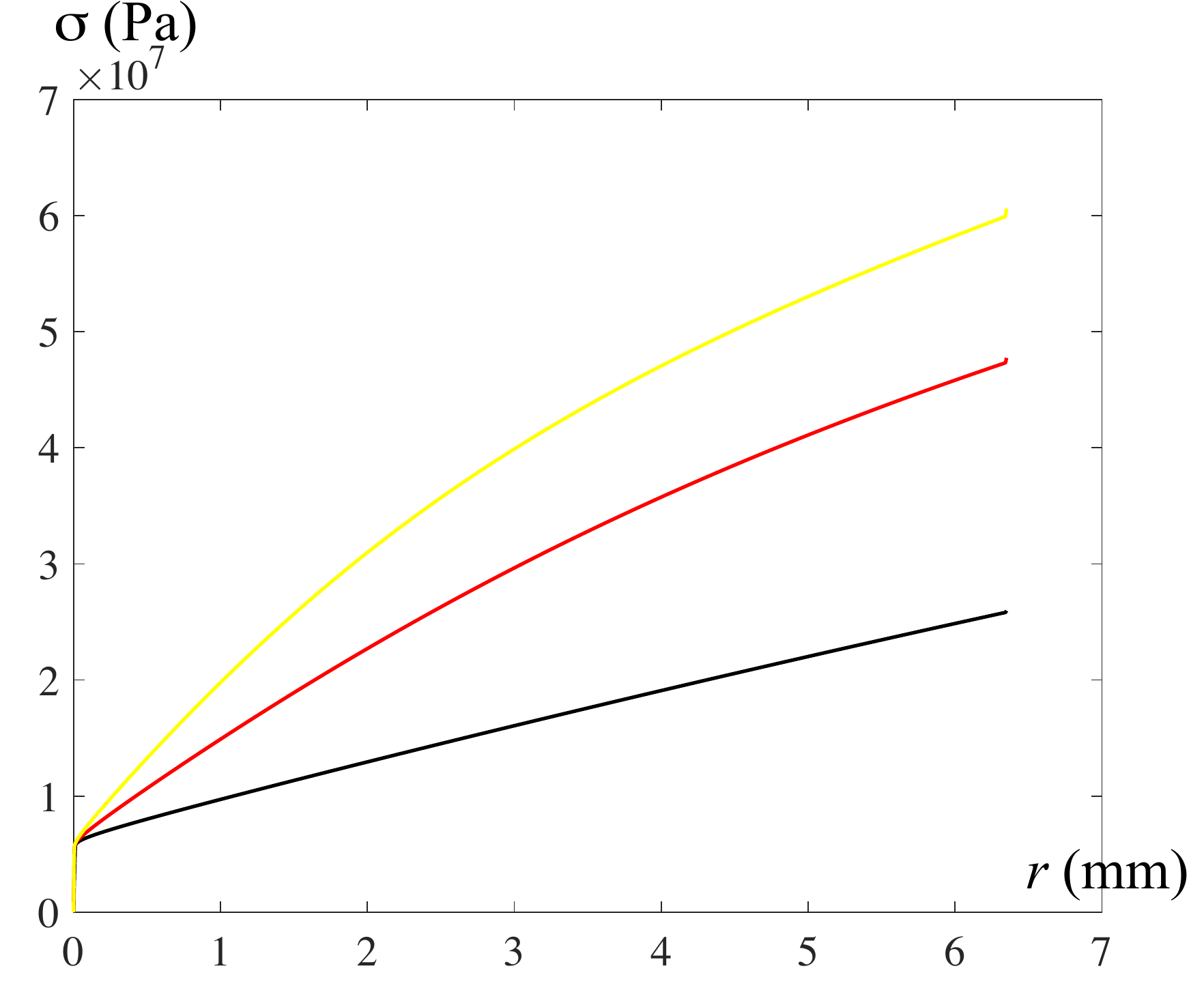}
	\caption{(Color online) Stress distribution $\sigma (r)$ at the twist rate $\dot{\phi}=0.25^\circ/$s and for room temperature: (i) $\phi =10^\circ$ (black), (ii) $\phi =30^\circ$ (red/dark gray), (iii) $\phi =50^\circ$ (yellow/light gray).}
	\label{Stress}
\end{figure}

In terms of the introduced quantities the governing equations read
\begin{eqnarray}
\frac{\partial \tilde{\tau}_Y}{\partial \phi} = \eta \left[\tilde{r}b - \frac{\tilde{q}(\tilde{\tau}_Y,\tilde{\rho})}{\tilde{\varpi }_0}\right], \label{tau} 
\\
\frac{\partial \tilde{\rho}}{\partial \phi} = \eta K_\rho \,\frac{\tilde{\tau}_Y\,\tilde{q}}{\tilde{\nu}(\theta,\tilde{\rho},\tilde{\varpi}_0r)^2\,\tilde{\varpi}_0}\, \Bigl[1 -\frac{\tilde{\rho}}{\tilde{\rho}_{ss}(\tilde{\chi})} \Bigr], 
\\ 
\frac{\partial \tilde{\chi }}{\partial \phi} = \eta K\,\frac{\tilde{\tau}_Y\,\tilde{q}}{\tilde{\varpi }_0}\,\Bigl( 1 -\frac{\tilde{\chi}}{\tilde{\chi}_0} \Bigr), 
\\ 
\tilde{r}\phi \eta b-\beta -\tilde{\tau}_B -\tilde{\tau}_Y =0, \label{force}
\end{eqnarray}
where $\tilde{\tau}_B$ is equal to
\begin{equation}
\label{backstress1}
-\frac{k_1\xi^2+(2k_0k_1-1)\xi+k_1k_0^2-2k_0}{4\pi(k_0+\xi)^2}(\beta_{,\tilde{r}\tilde{r}}+\beta_{,\tilde{r}}/\tilde{r}-\beta/\tilde{r}^2),
\end{equation}
with $\xi=|\beta_{,\tilde{r}}+\beta/\tilde{r}|$. To solve this system of partial differential equations subject to initial and boundary conditions numerically, we discretize the equations in the interval $(0< \tilde r <\tilde{R})$ by dividing it into $n$ sub-intervals of equal length $\Delta \tilde{r}=\tilde{R}/n$. The first and second spatial derivative of $\beta$ in equation (\ref{force}) are approximated by the finite differences
\begin{eqnarray}
\frac{\partial \beta}{\partial \tilde{r}}(\tilde{r}_i)=\frac{\beta_{i+1}-\beta_i}{\Delta \tilde{r}},
\\
\frac{\partial ^2\beta}{\partial \tilde{r}^2}(\tilde{r}_i)=\frac{\beta_{i+1}-2\beta_i+\beta_{i-1}}{(\Delta \tilde{r})^2},
\end{eqnarray}
where $\beta_i=\beta(\tilde{r}_i)$. For the end-point $\tilde{r}=\tilde{R}$ we introduce $\beta_{n+1}$ at a fictitious point $\tilde{r}_{n+1}=(n+1)\Delta \tilde{r}$ and find it from the discretized condition of vanishing density of excess dislocations
\begin{equation}
\label{bc}
\frac{\beta_{n+1}-\beta_n}{\Delta \tilde{r}}+\beta_n/\tilde{R}=0.
\end{equation}
Then it is possible again to discretize the first and second derivative of $\beta $ at $\tilde{r}=\tilde{R}$ and write the finite difference equation for $\beta $ at that point. In this way,  we reduce the four partial differential equations to a system of $4n$ ordinary differential-algebraic equations that will be solved by Matlab-ode15s. 
 
\begin{figure}[t]
	\centering
	\includegraphics[width=.45\textwidth]{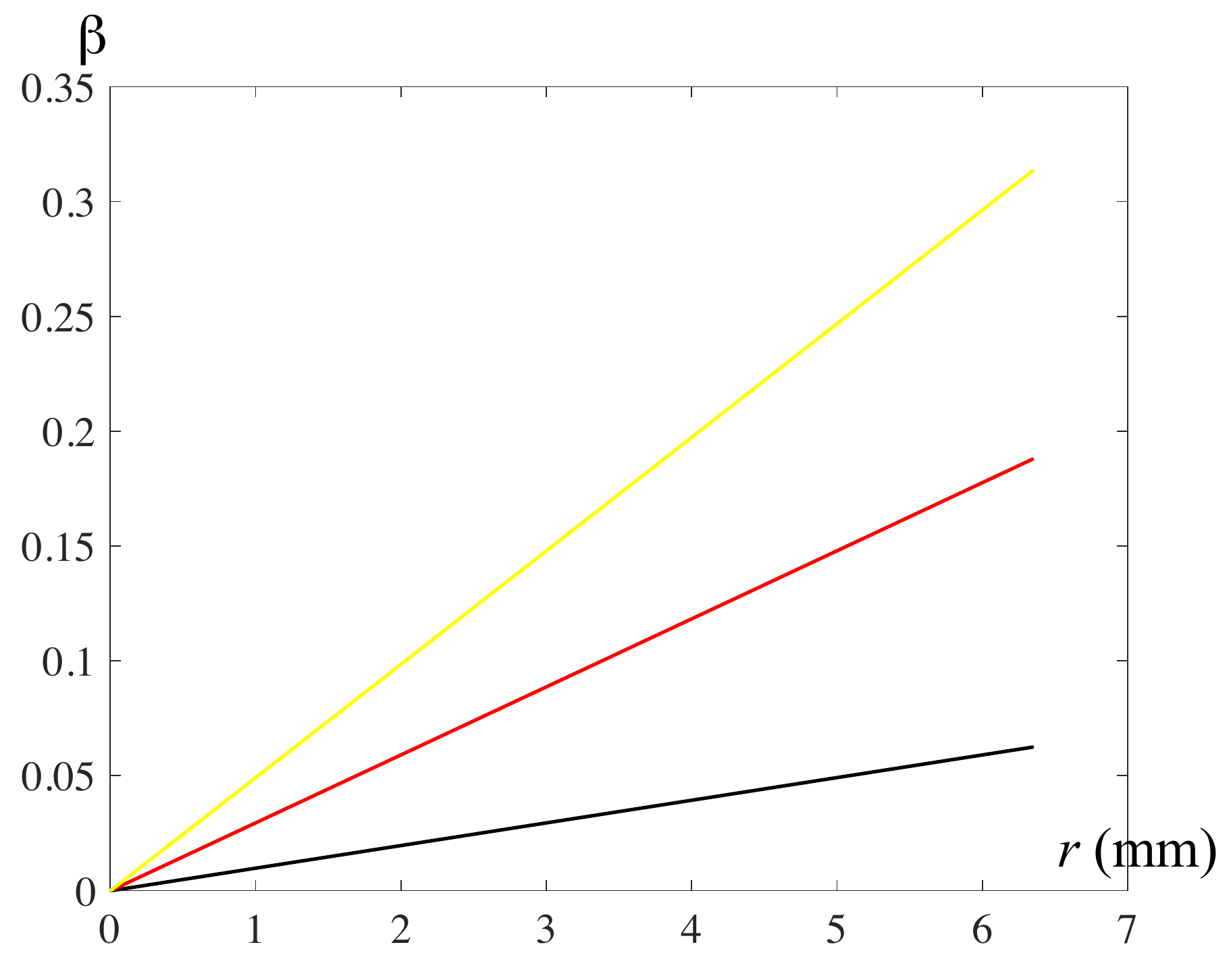}
	\caption{(Color online) Plastic distortion $\beta(r)$ at the twist rate $\dot{\phi}=0.25^\circ/$s and for room temperature: (i) $\phi =10^\circ$ (black), (ii) $\phi =30^\circ$ (red/dark gray), (iii) $\phi =50^\circ$ (yellow/light gray).}
	\label{Beta}
\end{figure}

After finding the solution we can compute the torque as function of the twist angle according to
\begin{equation}
\label{torque}
T = 2\pi \mu \int_0^{R} [r\phi \eta -\beta (r,\phi )] r^2 dr.
\end{equation}

\begin{figure}[t]
	\centering
	\includegraphics[width=.45\textwidth]{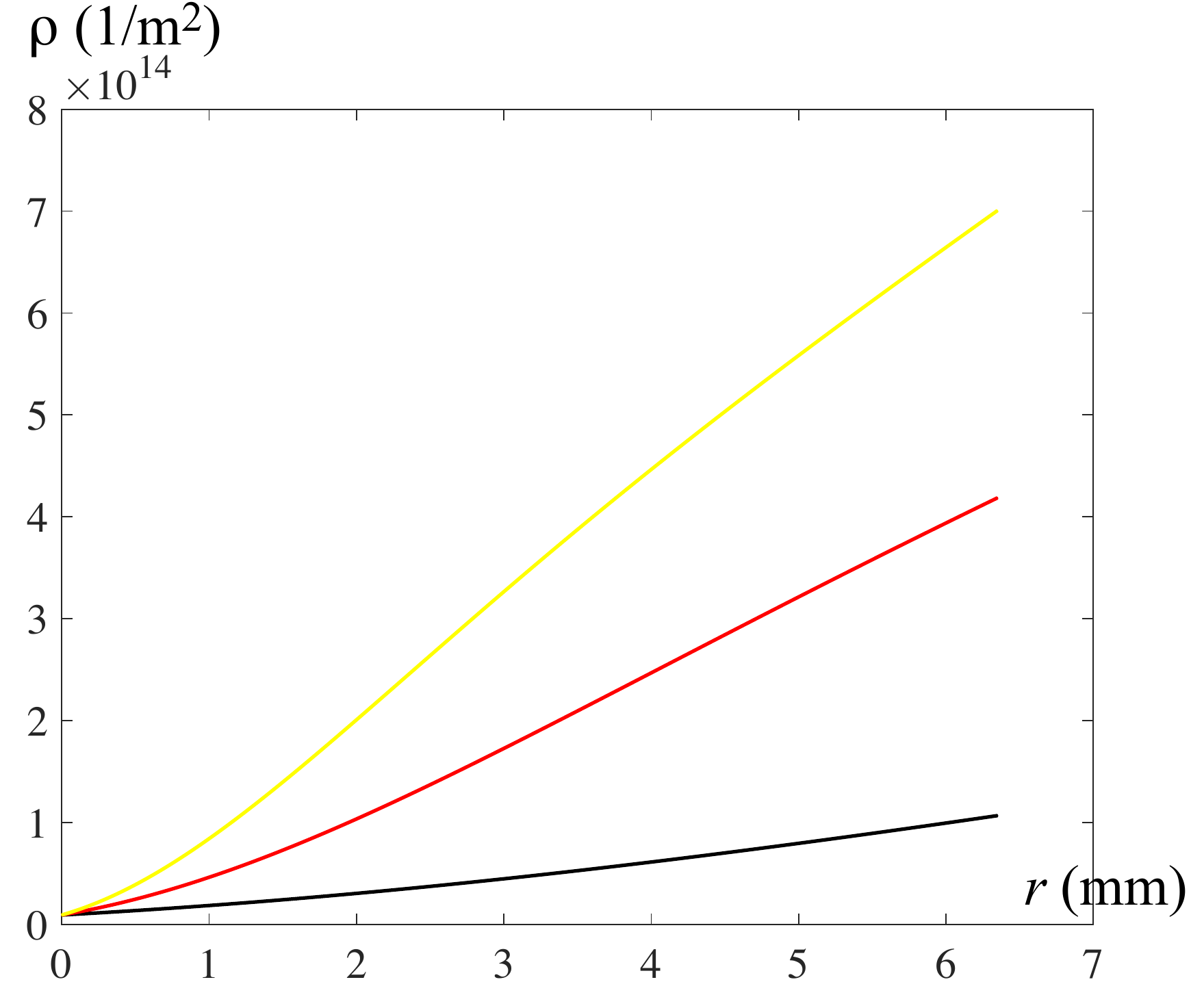}
	\caption{(Color online) Total density of dislocations $\rho(r)$ at the twist rate $\dot{\phi}=0.25^\circ/$s and for room temperature: (i) $\phi =10^\circ$ (black), (ii) $\phi =30^\circ$ (red/dark gray), (iii) $\phi =50^\circ$ (yellow/light gray).}
	\label{Densitytotal}
\end{figure}

\section{Parameter identification and numerical simulations}
\label{PI}

The experimental torque-twist curve for the single crystal copper bar provided in \cite{Horstemeyer02} along with our theoretical results based on the preceding  equations of motion are shown in Fig.~\ref{torquetwist}.  In this figure, the circles represent the experimental data for sample 1 in \cite{Horstemeyer02} while the solid curve is our theoretical simulation. The experimental data for sample 2 in that paper appear less reliable, especially at large twist angles, and are not analyzed here.

In order to compute the theoretical torque-twist curve, we need values for seven system-specific parameters and two initial conditions.  The seven basic parameters are the following: the activation temperature $T_P$, the stress ratio $s$, the steady-state scaled effective temperature $\tilde\chi_0$, the two dimensionless conversion factors $K_\rho$ and $K$, the two coefficients $k_0$, and $k_1$ defining the function $\tilde{\tau}_B$ in Eq.~(\ref{backstress1}). We also need initial values of the scaled dislocation density $\tilde\rho_i$ and the effective disorder temperature $\tilde\chi_i$; all of which are determined by the sample preparation. The other parameters required for numerical simulations but known from the experiment are: the ambient temperature $T=298$K, the shear modulus $\mu=48$GPa, the length $L=17.6$mm and radius $R=6.35$mm of the bar, the length of Burgers' vector $b=2.55$\AA, the twist rate $\dot{\phi}=0.25^\circ/$s, and consequently, $\tilde{\varpi}_0=0.2479\times 10^{-12}/$m. We take $a=10b$.

\begin{figure}[t]
	\centering
	\includegraphics[width=.45\textwidth]{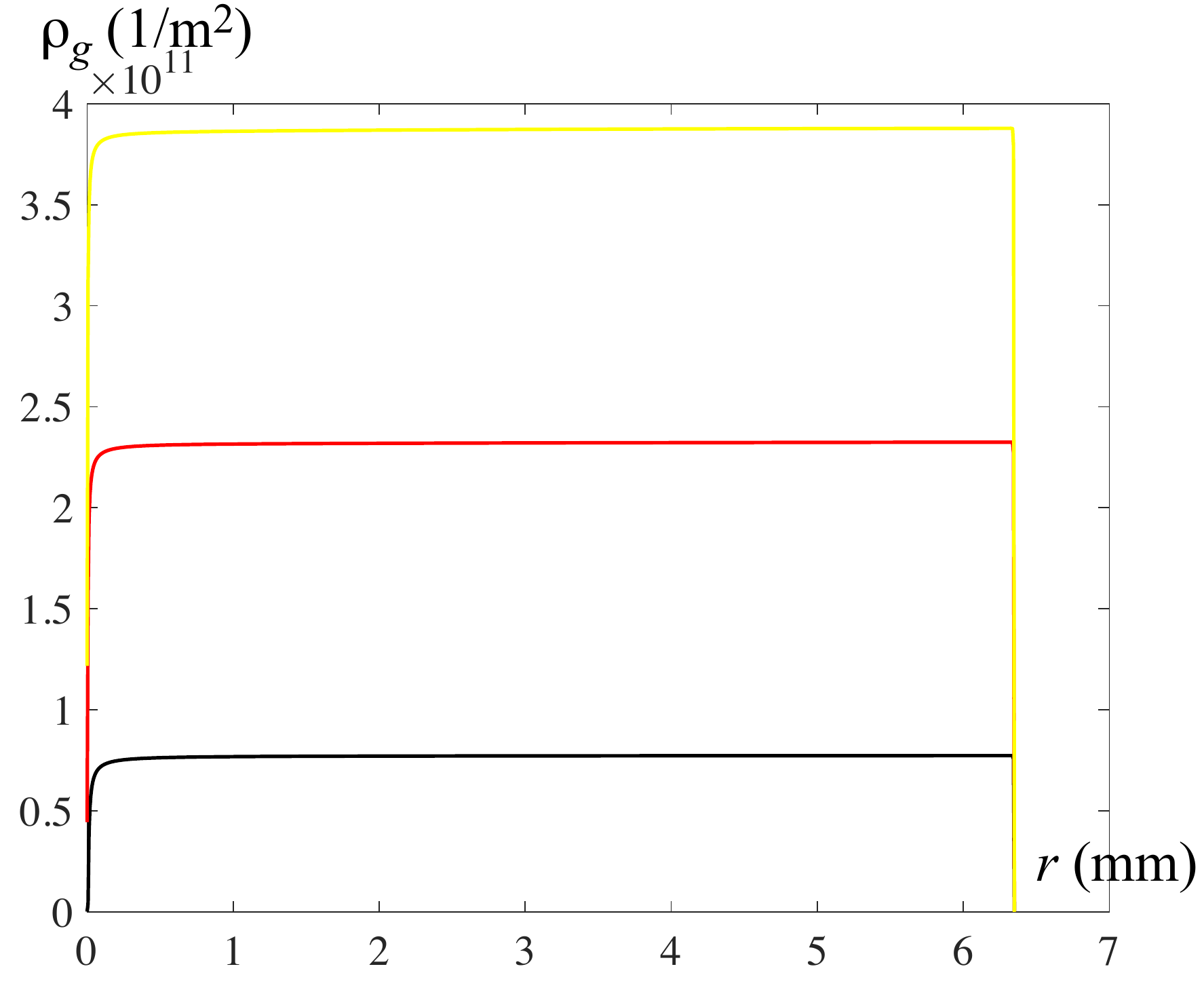}
	\caption{(Color online) Density of excess dislocations $\rho_g(r)$ at the twist rate $\dot{\phi}=0.25^\circ/$s and for room temperature: (i) $\phi =10^\circ$ (black), (ii) $\phi =30^\circ$ (red/dark gray), (iii) $\phi =50^\circ$ (yellow/light gray).}
	\label{DensityExcess}
\end{figure}
 
In earlier papers dealing with the uniform deformations \cite{LBL-10,JSL-15,JSL-16,JSL-17}, it was possible to begin evaluating the parameters by observing steady-state stresses $\sigma_{ss}$ at just a few strain rates $q_0$ and ambient temperatures $T_0 = T_P\,\tilde\theta_0$. Knowing $\sigma_{ss}$, $T_0$ and $q_0$ for three stress-strain curves, one could solve equation 
\begin{equation}
\label{qdef2}
\sigma = \sigma_T(\tilde\rho)\,\nu(\tilde\theta,\tilde\rho,q_0),
\end{equation} 
which is the inverse of Eq.~(\ref{qdef}) for $T_P$, $s$, and $\tilde\chi_0$, and check for consistency by looking at other steady-state situations. With that information, it was relatively easy to evaluate $K_\rho$ and $K$ by directly fitting the full stress-strain curves.  This strategy does not work here because the stress state of twisted bars is non-uniform. We may still have local steady-state stresses as function of the radius $r$, but it is impossible to extract this information from the experimental torque-twist curve. Furthermore, the similar parameters for copper found in \cite{LBL-10,JSL-15,JSL-16,JSL-17} cannot be used here, since we are dealing with screw dislocations having the energy barrier $T_P$ and other characteristics different from those identified in the above references.

To counter these difficulties, we have resorted to the large-scale least-squares analyses that we have used in \cite{Le17,LeTr17,Le18}. That is, we have solved the system of ordinary differential-algebraic equations (DAE) numerically, provided a set of material parameters is known. Based on this numerical solution we then computed the sum of the squares of the differences between our theoretical torque-twist curve and a large set of selected experimental points, and minimized this sum in the space of the unknown parameters. The DAE were solved numerically using the Matlab-ode15s, while the finding of least squares was realized with the Matlab-globalsearch. To keep the calculation time manageable and simultaneously ensure the accuracy, we have chosen $n=1000$ and the $\phi$-step equal to $\phi_*/7335$. We have found that the torque-twist curve for sample 1 taken from \cite{Horstemeyer02} can be fit with just a single set of system parameters.  These are: $T_P = 28911$\,K,\,$s = 0.0156,\,\chi_0= 0.243,\,K_\rho=49.2,\, K=379,\,k_0 = 4.31 \times 10^{-7}$,\,$k_1= 1.79\times 10^{8}$,\,$\tilde\rho_i=6\times 10^{-5}$, and $\tilde\chi_i=0.178$. The agreement between theory and experiment seems to us to be well within the bounds of experimental uncertainties.  Even the initial yielding transition appears to be described accurately by this theory. There is only one visible discrepancy: at large twist angles ($\phi>69^\circ$) the torques are slightly below those predicted by the theory.  Nothing about this result leads us to believe that there are relevant physical ingredients missing in the theory. 

\begin{figure}[t]
	\centering
	\includegraphics[width=.45\textwidth]{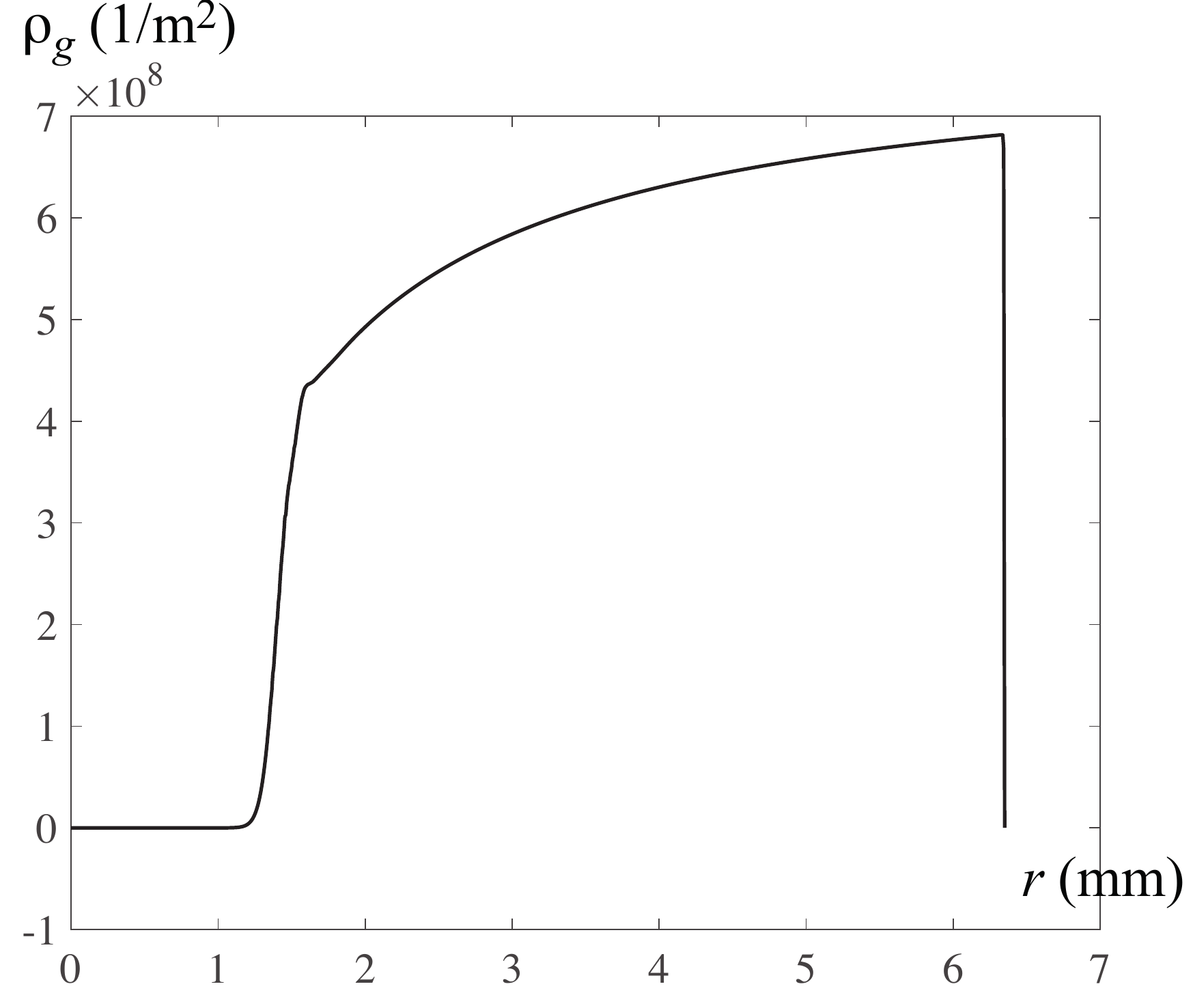}
	\caption{Density of excess dislocations $\rho_g(r)$ at the twist rate $\dot{\phi}=0.25^\circ/$s and for room temperature at $\phi =0.1^\circ$.}
	\label{DensityExcSmall}
\end{figure}
 
The results of numerical simulations for other quantities are shown in Figs.~\ref{Stress}-\ref{DensityExcSmall}. We plot in Fig.~\ref{Stress} the shear stress distribution $\tau =\mu (r\phi \eta -\beta )$ at three different twist angles $\phi=10^\circ$ (black), $\phi=30^\circ$ (red/dark gray), and $\phi=50^\circ$ (yellow/light gray). Contrary to the similar distribution obtained by the phenomenological theory of ideal plasticity, the stress in the plastic zone does not remain constant, but rises with increasing $r$ and reaches a maximum at $r=R$. This exhibits the isotropic hardening behavior due to the entanglement of dislocations. Fig.~\ref{Beta} shows the evolution of the plastic distortion $\beta (r)$ at the above three different twist angles. It can be seen that the plastic distortion is an increasing function of $r$ except very near the free boundary $r=R$. Since the latter attracts excess dislocations, $\beta (r)$ should decrease in this region to ensure equilibrium. However, due to the strong external stress field, the influence of this attraction can only be felt in a thin layer near the free boundary. In our approximate finite difference solution the decrease of $\beta (r)$ occurs between the end-point and the fictitious point and cannot be seen in Fig.~\ref{Beta}. Figs.~\ref{Densitytotal} and \ref{DensityExcess} present the densities of total and excess dislocations, respectively, at the above three different twist angles. Under the applied shear stress, the excess dislocations of the positive sign move to the center of the bar and pile up there. At large twist angles the distribution of excess dislocations over radius $r$ remains almost constant except near the center and the free boundary (cf. \cite{KaluzaLe11,LePiao16,Liu18}). 

\begin{figure}[t]
	\centering
	\includegraphics[width=.45\textwidth]{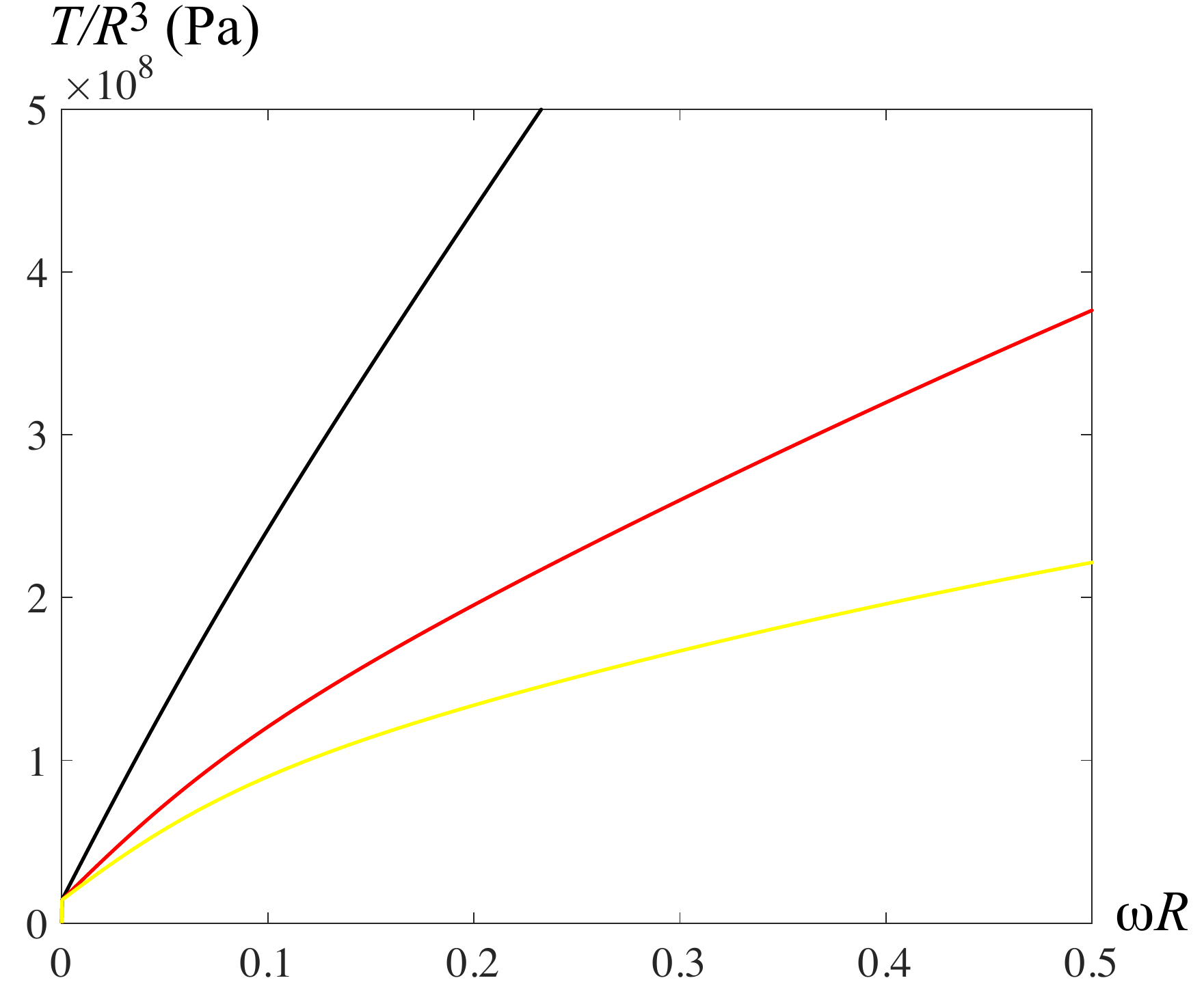}
	\caption{(Color online) Normalized torque $T/R^3$ (Pa) versus normalized twist $\omega R$ curves for bars with different radii at the maximum strain rate $\dot{\omega}R=10^{-3}/$s and for room temperature: (i) $R=25$ micron (black), (ii) $R=50$ micron (red/dark gray), (iii) $R=100$ micron (yellow/light gray).}
	\label{Size}
\end{figure}

To understand the mechanism of formation of excess dislocations, we plot in Fig.~\ref{DensityExcSmall}  distribution $\rho_g(r)$ at a small twist angle $\phi=0.1^\circ$. Since the flow stress at this twist angle exceeds the Taylor stress, redundant dislocations in the form of dislocation dipoles begin to dissolve according to the kinetics of thermally activated dislocation depinning \cite{LBL-10,JSL-15,JSL-16,JSL-17,JSL-17a}. Under the applied shear stress, positive dislocations then move towards the center and negative dislocations towards the boundary. For the dissolved dislocation dipoles within the sample and far from the free boundary, these freely moving dislocations are soon trapped by dislocations of the opposite sign. But the dislocation dipoles near the free boundary behave differently. Now the positive dislocations move inwards and become excess dislocations, while the negative dislocations leave the sample and become image dislocations. At small angles of twist, the applied shear stress near the center is still small and cannot move dislocations. Therefore, excess dislocations occupy an outer ring, as can be seen in Fig.~\ref{DensityExcSmall}. As the angle of twist increases, the shear stress increases as well, and when it becomes large enough, it can drive these excess dislocations to the center and they pile up there. Thus, we can say that the dissolution of dipoles near the free boundary results in excess dislocations of positive sign. They then move to the center and pile up there, increasing kinematic hardening.

\begin{figure}[t]
	\centering
	\includegraphics[width=.45\textwidth]{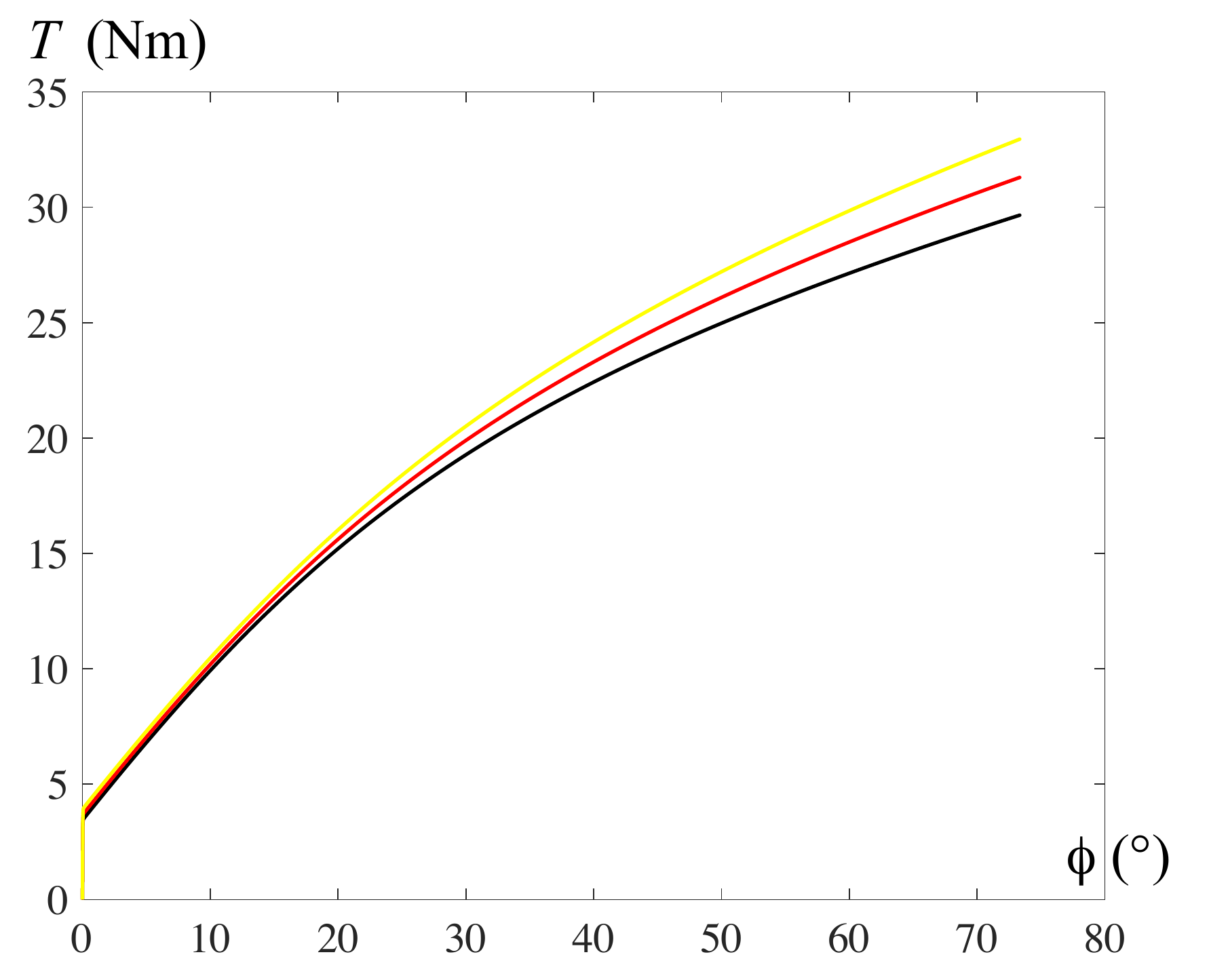}
	\caption{(Color online) The torque-twist curves for the bars twisted at different twist rates and for room temperature: (i) $\dot{\phi}=0.25^\circ/$s (black), (ii) $\dot{\phi}=2.5^\circ/$s (red/dark gray), (iii) $\dot{\phi}=25^\circ/$s (yellow/light gray).}
	\label{StrainRate}
\end{figure}
 
It is interesting to examine the influence of the size of the sample on the torque-twist curve. Fig.~\ref{Size} shows the three normalized torque $T/R^3$ (measured in Pa) versus normalized twist $\omega R$ curves for three bars with different radii $R=25$ micron (black), $R=50$ micron (red/dark gray), and $R = 100$ micron (yellow/light gray). We choose the maximum strain rate $\dot{\omega}R=10^{-3}$/s, while all other parameters are left unchanged. We see that the size strongly influences the slope of the hardening curve, since the accumulated excess dislocations pile up against the center leading to a stronger kinematic hardening for the smaller sample than for the larger one (smaller is stronger). The yielding transition, on the other hand, is almost independent of the radius. This can be explained by the fact that at the onset of yielding transition practically no excess dislocations occur, so that the kinematic hardening is not yet noticeable.

Another important question is how strongly the twist rate affects the torque-twist curve. Fig.~\ref{StrainRate} shows the three torque-twist curves for three samples loaded at three different twist rates $\dot{\phi}=0.25^\circ/$s (black), $\dot{\phi}=2.5^\circ/$s (red/dark gray), and $\dot{\phi}=25^\circ/$s (yellow/light gray). The radius of the samples is $R=6.35$mm, while all other parameters remain unchanged. We see that the twist rate mainly affects isotropic hardening: the higher the twist rate, the higher the slope of the torque-twist curve. The kinematic hardening is not affected by the change of the twist rate. The reason for this is that the kinematic hardening due to the excess dislocations is much less sensitive to the change in strain (twist) rate. 

\section{Concluding Remarks}
\label{CONCLUSIONS}
  
Overall, these results seem to us to be quite satisfactory.  Note that we now use thermodynamic dislocation theory for non-uniform deformations not just to test its validity but also as a tool for discovering properties of structural materials. For example, we could find the mechanism of forming excess dislocations based on the dissolution of dislocation dipoles near the free boundary of the bar and predict their distribution. One of the main reasons for the success of this theory -- as has been emphasized here and in earlier papers -- is the extreme sensitivity of the plastic strain rate to small changes in the temperature or the stress. Another reason for its success is the inclusion of the excess dislocations in the theory, which leads to size-dependent kinematic hardening. Here, in our opinion, the incompatible plastic distortion is the natural variable that keeps the memory of excess dislocations. It cannot enter the free energy, but the curl of this quantity should enter the free energy causing the back stress. In this way the theory differs substantially from the phenomenological plasticity that introduces the back stress along with an assumed constitutive equation to fit the stress strain curves with kinematic hardening. On the contrary, our theory allows us to find the back stress from the first principle calculation of the free energy of dislocated crystals.

The results obtained show the principal applicability of TDT to non-uniform plastic deformations. As far as the size effect is concerned, we could not find reliable experimental data for single crystal copper under torsion at different bar radii, in contrast to polycrystalline copper under torsion \cite{Fleck94}. However, the proposed theory may serve as a useful guide for the future experimental investigation of the torsion of single crystal bars in several directions: (i) the torque-twist curves at load reversals and the analog of the Bauschinger effect, (ii) the size effect, (iii) the sensitivity of the torque-twist curves to the twist rate and temperature, et cetera. The identification of material parameters for polycrystalline copper under torsion and the comparison with experiments in \cite{Fleck94} will be addressed in our forthcoming paper.

\begin{acknowledgments}

Y. Piao and T.M. Tran acknowledge support from the Chinese and Vietnamese Government Scholarship Program, respectively. K.C. Le is grateful to J.S. Langer for helpful discussions.

\end{acknowledgments}


\begin{thebibliography}{99}
 

\bibitem{LBL-10} J.S. Langer, E. Bouchbinder and T. Lookman, Acta Mater. {\bf 58}, 3718 (2010).

\bibitem{JSL-15} J.S. Langer, Phys. Rev. E {\bf 92}, 032125 (2015).

\bibitem{JSL-16} J.S. Langer, Phys. Rev. E {\bf 94}, 063004 (2016).

\bibitem{JSL-17} J.S. Langer, Phys. Rev. E. {\bf 95}, 013004 (2017).

\bibitem{JSL-17a} J.S. Langer, Phys. Rev. E. {\bf 95}, 033004 (2017).

\bibitem{Le17} K.C. Le, T.M. Tran and J.S. Langer, Phys. Rev. E. {\bf 96}, 013004 (2017).

\bibitem{Le18} K.C. Le, T.M. Tran and J.S. Langer, Scripta Mater. {\bf 149}, 62 (2018).

\bibitem{Le18a} K.C. Le, J. Mech. Phys. Solids {\bf 111}, 157 (2018).

\bibitem{LP18} K.C. Le and Y. Piao, arXiv:1801.05304 (2018).

\bibitem{LeTr18} K.C. Le, T.M. Tran, Phys. Rev. E. {\bf 97}, 043002 (2018).

\bibitem{Nye53} J.F. Nye, Acta Metall. {\bf 1}, 153 (1953).

\bibitem{Fleck94} N.A. Fleck et al., Acta Metall. Mater. {\bf 42}, 475 (1994). 

\bibitem{Stoelken98} J.S. St\"olken and A.G. Evans, Acta Mater. {\bf 46}, 5109 (1998). 

\bibitem{Horstemeyer02} Horstemeyer et al., Trans. ASME, {\bf 124}, 322 (2002).

\bibitem{LeTr17} K.C. Le, T.M. Tran, Int. J.  Eng. Sci. {\bf 119}, 50 (2017).

\bibitem{Kroener1992}
E. Kr{\"o}ner, GAMM-Mitteilungen {\bf 15}, 104 (1992). 

\bibitem{VB17} V.L. Berdichevsky, Int. J.  Eng. Sci. {\bf 116}, 74 (2017). 

\bibitem{KaluzaLe11} M. Kaluza, K.C. Le, Int. J. Plasticity {\bf 27}, 460 (2011).

\bibitem{LePiao16} K.C. Le, Y. Piao, Int. J. Plasticity {\bf 83}, 110 (2016).

\bibitem{Liu18} D. Liu et al., Acta Mater. {\bf 150}, 213 (2018).

\end{thebibliography}
\end{document}